\documentclass[aps,pra,twocolumn,superscriptaddress]{revtex4}
\usepackage{graphicx}
\usepackage{amssymb}
\usepackage{subfigure}
\usepackage{amsmath}
\usepackage{amsfonts}
\usepackage{bm}
\usepackage{color}
\newcommand{\ket}[1]{\ensuremath{\left|{#1}\right\rangle}}
\newcommand{\bra}[1]{\ensuremath{\left\langle{#1}\right |}}

\begin{document}


\title{Maximizing Post-selected Quantum Correlations from Classical Interference in a Multi-core Fiber Beamsplitter}

\author{J.~Cari\~{n}e}
\affiliation{Departamento de Ingenier\'ia El\'ectrica, Universidad Cat\'olica de la Sant\'isima Concepci\'on, Alonso de Ribera 2850, Concepci\'on, Chile}
\affiliation{ANID – Millennium Science Initiative Program – Millennium Institute for Research in Optics, Universidad de Concepci\'on, 160-C Concepci\'on, Chile}

\author{M. Asan-Srain}
\affiliation{Departamento de F\'{\i}sica, Universidad de Concepci\'on, 160-C Concepci\'on, Chile}
\author{G. Lima}
\affiliation{Departamento de F\'{\i}sica, Universidad de Concepci\'on, 160-C Concepci\'on, Chile}
\affiliation{ANID – Millennium Science Initiative Program – Millennium Institute for Research in Optics, Universidad de Concepci\'on, 160-C Concepci\'on, Chile}
\author{S. P. Walborn}
\affiliation{Departamento de F\'{\i}sica, Universidad de Concepci\'on, 160-C Concepci\'on, Chile}
\affiliation{ANID – Millennium Science Initiative Program – Millennium Institute for Research in Optics, Universidad de Concepci\'on, 160-C Concepci\'on, Chile}
\begin{abstract}
Fourth-order interference is an information processing primitive for photonic quantum technologies.  When used in conjunction with post-selection, it forms the basis of photonic controlled logic gates, entangling measurements, and can be used to produce quantum correlations.  Here, using classical weak coherent states as inputs, we study fourth-order interference in novel $4 \times 4$ multi-port beam splitters built within multi-core optical fibers.   Using two mutually incoherent weak laser pulses as inputs, we observe high-quality fourth order interference between photons from different cores, as well as self-interference of a two-photon wavepacket.  In addition, we show that quantum correlations, in the form of quantum discord, can be maximized by controlling the intensity ratio between the two input weak coherent states.  This should allow for the exploitation of quantum correlations in future telecommunication networks.     
\end{abstract}



\maketitle
 \section{Introduction}
 Quantum information promises to revolutionize the way in which information is transmitted, processed and stored, allowing for novel paradigms such as quantum cryptography and quantum computing.  A key element in this effort is the need to send quantum information from one place to another.  In this regard, quantum communication will most likely need to employ the same technological infrastructure as classical telecommunications.  For example, a main goal in telecommunications is to increase the transmission capacity of optical channels.  This has led to a number of novel encoding schemes as well as new technologies.  Currently, data rates are nearing the physical limits that are possible in single-mode optical fibers, known as the ``capacity crunch".  One promising and exciting solution to this problem is the use of multi-core optical fibers (MCF), composed of multiple fiber cores within the same cladding \cite{xavier20}.
 \par
 Multi-core optical fibers might have an even bigger impact on quantum information protocols.  For one, the relative phase fluctations between quantum states corresponding to different cores in the same cladding is much less than for different single-mode fibers \cite{lio20}.  This has led to a number of experiments involving quantum systems with dimension greater than two \cite{Canas17,ding17,carine20,gomez20,taddei20}.  However, a complete toolbox for the manipulation of photonic quantum information encoded in MCFs is still lacking.  One important element, fiber-imbedded multi-core beam splitters (MCF-BS), has been recently reported in the context of single-photon experiments  \cite{carine20}.   These are multi-port interference devices that coherently combine light from more than two input fiber cores, without removing the light from the cladding material.  They are thus multi-port interference devices that can be used to decrease the optical depth of linear optical circuits \cite{puma,saygin20}.    
 \par        
One important application of optical beam splitters is the realization of fourth-order interference \cite{hom87}.  In the quantum regime, this is used to build photon controlled-logic gates \cite{klm01,ralph02}, and in projection onto entangled states \cite{mattle96,silva13,aguilar19,piera20}.  Moreover, fourth-order interference can produce quantum correlations when the quantum state is post-selected in the number basis.  We note that this was the first source of polarization-entangled photon pairs using photons from spontaneous parametric down-conversion \cite{shih88,ou88}.         
\par
 Here we investigate multi-photon interference in a MCF-BS using two independent weak coherent states (WCS) at telecom wavelengths.  There has been much recent interest in this scenario since the development of measurement-device independent quantum key distribution \cite{lo12,wang17}. This is a classical fourth-order interference effect \cite{silva15,hong17}, where photons from two input WCSs show bunching and anti-bunching behavior when photon pairs are detected in coincidence at different output ports.  A similar approach reported in  Ref. \cite{choi17} for equal intensity WCSs showed that the post-selected output state, though not entangled, could demonstrate non-classical correlations in the form of quantum discord \cite{henderson01,ollivier02}.     Recent studies have shown that quantum discord plays an important role in quantum tasks such as quantum computing \cite{datta08}, remote state preparation \cite{dakic12}, quantum illumination and metrology \cite{girolami14,weedbrook16}, quantum cryptography \cite{pirandola14}, and the quantum-classical transition \cite{mazzola10,cornelio12}.  Moreover, a number of interesting dynamics have been explored \cite{maziero09,fanchini10,shi10,auccaise11,werlang10,maziero10,lanyon13}. A more comprehensive review of discord and its relation to quantum phenomena and protocols can be found in Refs. \cite{celeri11,modi12,bera17,hu18}. 
 \par
We explore the creation of quantum correlations in a four-port fiber beam-splitter (4CF-BS) device both theoretically and experimentally.  In section \ref{sec:expdes}, we present our experimental setup and the characterization of two different types of fourth-order interference.  The first involves interference between photons originating from mutually incoherent WCSs.  The second involves interference between two-photon wavepackets produced in the output of the 4CF-BS device.   We observe high-quality interference in both cases.  In section \ref{sec:discord}, we then consider input laser pulses with variable intensities, and show that the post-selected quantum correlations between photonic qubits can be maximized as a function of the intensity mismatch ratio.   We evaluate the quantum correlations using the geometric discord \cite{dakic10}, and determine the optimal mismatch ratio.    
\par
  
 \section{Experimental Description}
 \label{sec:expdes}
\begin{figure}
 \includegraphics[width=8cm]{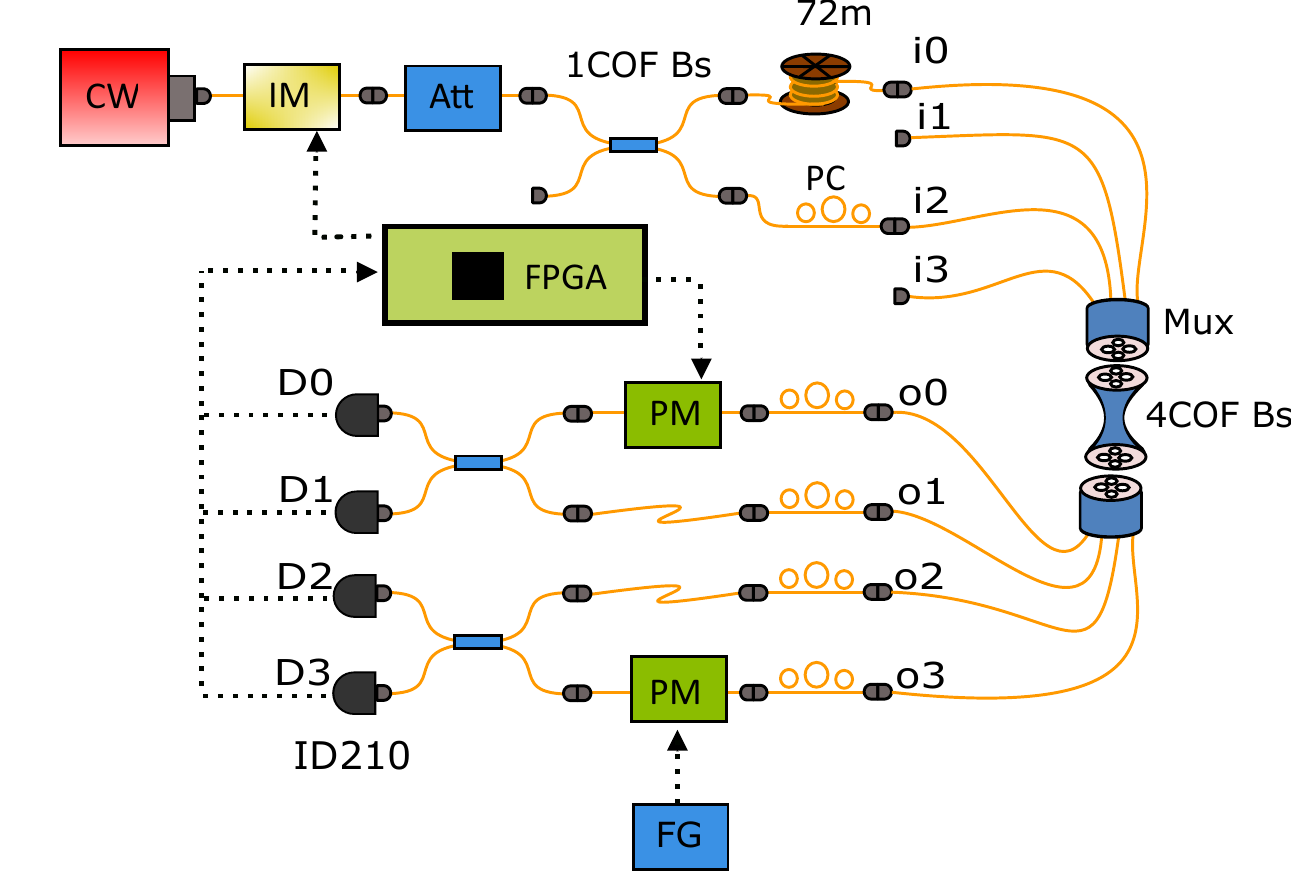}
\caption{Experimental setup. Weak coherent pulses are interfered incoherently on a four-core fiber beam splitter (4CF-BS). CW: continuous wave laser, IM: intensity modulator, Att: attenuator, BS: single-mode fiber $2 \times 2$ beam splitter, PC: polarizer controller, MUX: fiber multiplexer for single-core SMF fiber to multicore fiber, 4CF-BS: four-core beam splitter, PM: phase modulator, FG: function generator, ID210: triggered single photon detector $D_j$, FPGA: field programmable gate array device.}
\label{fig:setup}
\end{figure}

The experimental setup is shown in figure \ref{fig:setup}.   The amplitude of a  fiber-coupled continuous wave laser with a  wavelength $\lambda =1550$ nm is amplitude modulated using a fiber-coupled Lithium Niobate intensity modulator (IM). By properly toggling the relative phase between 0 and $\pi$, the output is a train of almost gaussian pulses of width 5ns at a repetition rate of 2.78Mhz.  These pulses are attenuated and then sent into a 50/50 fiber BS, resulting in two WCSs.  These WCSs are sent to inputs 1 and 3 of the 4CF-BS using a demultiplexer device (DEMUX), used to couple single mode fibers (SMF) to different cores of a MCF.  However, one of the WCSs is first sent through 72m of optical fiber, so that the WCS pulses that overlap temporally at the 4CF-BS are mutually incoherent.   This relative delay is much longer than the coherence length of the laser, so that there is no second-order (single-photon) interference between the overlapping pulses.  However, fourth-order (two-photon) interference can occur \cite{silva13,silva15,hong17,choi17}. The 4CF BS outputs are connected to two independent two-arm interferometers, where each new $2 \times 2$ interferometer is phase controlled with a fiber-coupled lithium niobate phase modulator (PM).
 Then, fourth order interference is measured from the detections of four triggered single photon detectors (SPD). In our setup, IdQuantique ID210 detectors were used.
The generation of pulses with the IM, in the WCSs source, and the measurement using the PMs and the SPD in the measurement stage is synchronized by a field programmable gate array (FPGA), which also records individual photon counts and coincidence counts in a  coincidence window of 1 ns wide. In order to evaluate the variation in interference, a simple electrical signal generating device (or function generator, FG) is incorporated (see Fig \ref{fig:setup}).
\subsection{$4\times4$ Multi-port fiber beamsplitter}
 We use the $4 \times4$ multicore fiber beam splitter (4CF-BS).  Fabrication and characterization of the device has been previously reported in Ref. \cite{carine20}.  The transfer matrix relating the input and output modes is described by 
 \begin{equation}
\mathbf{B}_4 = \frac{1}{2} \left (
  \begin{array}{cccc}
  1 & 1 & 1 &  1\\
  1 & 1 & -1 & -1 \\
   1 & -1& 1& -1 \\
 1 & -1 & - 1 & 1 
  \end{array}
\right ).
\label{eq:MCF-BS}
\end{equation}
We can denote the input and output creation operators using the vectors $\vec{a}^\dagger_{in/out} = (\hat{a}_0^\dagger,\hat{a}_1^\dagger,\hat{a}_2^\dagger,\hat{a}_3^\dagger)^T$, so that $\vec{a}^\dagger_{out} = \mathbf{B}_4 \vec{a}^\dagger_{in}$. 
 \par
 \subsection{Input and Output state with Post-Selection}
 \label{sec:PSstate}
A coherent state can be written in the Fock basis as
 \begin{equation}
 \ket{\eta} = e^{-|\eta|^2/2} \sum\limits_{n=0}^\infty \frac{\eta^n}{\sqrt{n!}} \ket{n}.
 \end{equation}
The Fock states are defined as $\ket{n} = (\hat{a}^\dagger)^n \ket{0}/\sqrt{n!}$, where $\hat{a}^\dagger$ is the photonic creation operator and $\ket{0}$ is the vacuum state.
 Let us consider two input weak coherent states with amplitudes $|\eta| << 1$ and $|\eta^\prime| << 1$  that are mutually incoherent, input into modes 1 and 3 of $\mathbf{B}_4$.    Ignoring terms containing more than two photons, the input state can be written as
  \begin{equation}
 \rho_{in} = N [p_{2}p^\prime_{0}\ket{2,0} \bra{2,0} + p_{0}p^\prime_{2}\ket{0,2} \bra{0,2}) +p_{1}p_1^\prime \ket{1,1} \bra{1,1}],
 \end{equation}
 where $p_n =  |\eta|^{2n} \exp(-|\eta|^2) /n!$ is the probability that the coherent state contains $n$ photons, $N$ is a normalization constant, and the bras and kets refer to input modes 1,3. Using the fact that $p_2p_0=p_1^2/2$, we can write
   \begin{equation}
 \rho_{in} = C  (\ket{2,0} \bra{2,0} + \gamma^2 \ket{0,2} \bra{0,2} +2 \gamma \ket{1,1} \bra{1,1}).
 \label{eq:rho_in}
 \end{equation}
 where $\gamma=\eta/\eta^\prime$ is the ratio between mean photon numbers, and $C$ is a normalization constant. 
 Since state  \eqref{eq:rho_in} is a convex sum of different Fock product states, we can transform each component using matrix \eqref{eq:MCF-BS}, and sum the results.  The two photon state $\ket{2,0}=(\hat{a}_1^\dagger)^2 \ket{0,0}/\sqrt{2}$.  The field operators transform as
    \begin{equation}
(\hat{a}_1^\dagger)^2 \rightarrow \frac{1}{4}(\hat{a}_0^\dagger+\hat{a}_1^\dagger-\hat{a}_2^\dagger-\hat{a}_3^\dagger )^2. 
 \end{equation}
 For the state $\ket{0,2}=(\hat{a}_3^\dagger)^2 \ket{0,0}/\sqrt{2}$, the field operator 
  \begin{equation}
(\hat{a}_3^\dagger)^2  \rightarrow \frac{1}{4}(\hat{a}_0^\dagger-\hat{a}_1^\dagger-\hat{a}_2^\dagger+\hat{a}_3^\dagger )^2.
 \end{equation}
 Finally, for the state $\ket{1,1}=\hat{a}_1^\dagger \hat{a}_3^\dagger \ket{0,0}$, we have the corresponding transformation 
  \begin{equation}
\hat{a}_1^\dagger \hat{a}_3^\dagger \rightarrow \frac{1}{4}\left( \hat{a}_0^{\dagger 2} -\hat{a}_1^{\dagger 2} +\hat{a}_2^{\dagger 2}-\hat{a}_3^{\dagger 2} - 2 \hat{a}^\dagger_0 \hat{a}^\dagger_2 + 2 \hat{a}^\dagger_1 \hat{a}^\dagger_3\right). 
 \end{equation}
 Here the absence of the terms $\hat{a}^\dagger_0 \hat{a}^\dagger_1$, $\hat{a}^\dagger_0 \hat{a}^\dagger_3$, $\hat{a}^\dagger_1 \hat{a}^\dagger_2$ and $\hat{a}^\dagger_2 \hat{a}^\dagger_3$ is due to two-photon interference.  That is, when the two photons are indistinguishable, these terms vanish due to destructive interference.    
 \par
  Using the above transformations in the input state \eqref{eq:rho_in}, The output state is given by a $10 \times 10$ density matrix, which can be written in the basis of possible two photon states in the four output cores of the 4CF-BS. For example, state $\ket{2000}$ correspond to two photons in output mode 0, state $\ket{0020}$ to two photons in output mode 2, and state $\ket{1010}$ to one photon in mode 0 and one photon in mode 2.  The complete density matrix written in this number basis is provided in the appendix.  
 \par
 \subsection{Fourth order interference}

 \subsubsection{Bunching and Anti-bunching}
  \label{sec:hom}
 Two photon interference between the WCS inputs can be observed by registering two-photon coincidence events at detectors connected to different output cores of the 4CF-BS.  Let us consider equal intensity WCSs, so that $\gamma=1$. The probability $P_{jk}$ to detect one photon in output core $j$ and the other in output mode $k$ ($j < k$) is given by the diagonal elements of the lower $6\times 6$ block of the density matrix in Eq. \eqref{eq:rhototequal}.   These probabilities are  $P_{01}=P_{03}=P_{12}=P_{23}=1/16$, and $P_{02}=P_{13}=3/16$.  To see that these values correspond to two-photon interference, we can consider the case of distinguishable photons, in which all of the 16 two-photon events are equally likely.  Since there are two events that result in one photon in mode $j$ and another in mode $k$ (e.g. $P_{jk}$ results from photon 1 going to $j$ and photon 2 to $k$, or vice versa), these output probabilities would be equal to $1/8$.  Thus, comparing the indistinguishable case with the distinguishable one, events with $P_{jk}=1/16$ correspond to destructive interference, and those with $P_{jk}=3/16$  to constructive interference.  The predicted interference visibility for WCSs is given by $\mathcal{V}_{WCS}=(3/16-1/16)/(3/16+1/16)=1/2$.  This is the classical limit for fourth-order interference, compared to the unit visibility which is achievable in principle with a pair of input Fock states \cite{hom87}.    
 \par
\begin{figure}
\includegraphics[width=8cm]{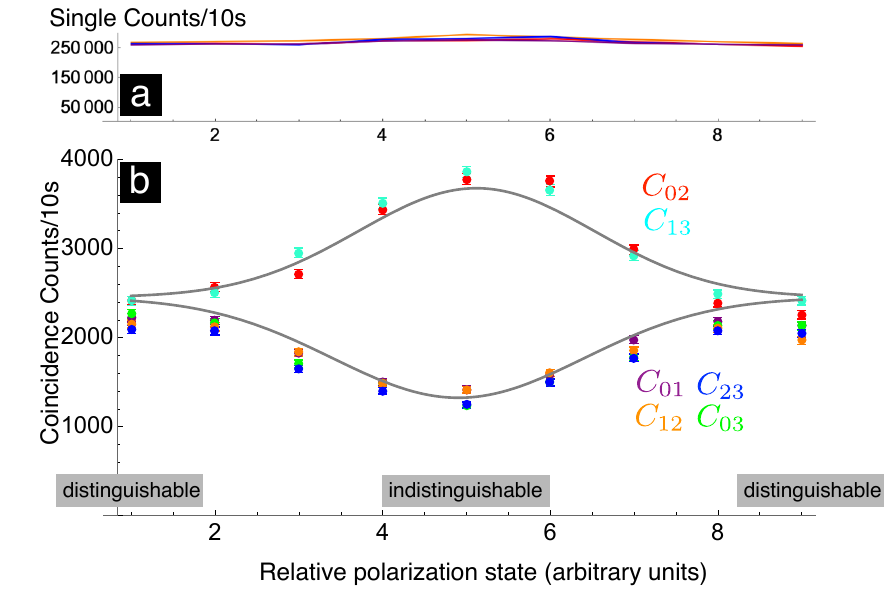}
\caption{Coincidence counts $C_{jk}$ between detectors $j$ and $k$ showing Interference of independent WCSs at a 4CF beam splitter.  Constructive and destructive fourth-order interference can be observed in the coincidence counts between pairs of detectors as a function of the orthogonality of the polarization states, while the single counts (top) remain unchanged.  Error bars correspond to Poissonian count statistics. The solid grey curves represent gaussian curve fits to experimental data, and are intended merely as a guide for the eye. }
\label{fig:HOM}
\end{figure}

Fourth order interference was explored by adjusting the polarization state of one of the WCSs by rotating one paddle of a fiber polarization controller, so that we can change continuously between distinguishable and indistinguishable photons.  
 As a function of a distinguishability parameter $\theta$ related to the overlap between polarization states, the output probability is 
 \begin{equation}
P_{jk}=\frac{1}{8}\left(1 \pm \frac{1}{2} f(\theta) \right ),
\end{equation} 
  where the plus (minus) sign refers to constructive (destructive) interference and $f(\theta)$ describes the polarization rotation implemented. In these measurements, the the output cores of the 4CF-BS were connected directly to the single-photon detectors using a DEMUX device. Coincidence counts $C_{jk}$ between detectors $j$ and $k$ were recorded as a function of $\theta$ and are shown in  Fig.  \ref{fig:HOM}.   When the polarization states are orthogonal, input photons from different pulses are distinguishable, and all combinations of coincidence counts have the same count rates.  When the polarization states are parallel, fourth-order interference occurs, resulting in an increase in coincidence counts $C_{02}$ and $C_{13}$, and  a suppression of the coincidence counts at other detector pairs.   Shown also are the single counts at each detector, which remain constant as the polarization is varied. This confirms that the increase or supression in coincidence counts is a fourth-order interference phenomenon.   Using the average values at the interference maximum and minimum, we obtained a visibility of $\mathcal{V}_{WCS}=0.48\pm0.02$, in agreement with the predicted value of $1/2$.       
 \subsubsection{Two photon wave packets}
As figure \ref{fig:HOM} shows, photon coalescence occurs for four combinations of coincidence outputs.  For these events,  the suppression of coincidence counts indicates that it is more likely to have both photons exiting the 4CF-BS in the same output core.  Moreover, the quantum state given by matrix \eqref{eq:rhototequal} contains coherences between these two-photon states.  For example, if we  isolate events $\ket{20}_{01}$, $\ket{11}_{01}$, and $\ket{02}_{01}$  involving two photons in output fibers 0 and 1, we have  the normalized density operator
 \begin{equation}
\rho_{2A} = 
\frac{1}{8} \left(
\begin{matrix}
3 & 0 & -1 \\
0 & 2 & 0 \\
-1 & 0  & 3 \\
\end{matrix}
\right).
 \end{equation} 
 We can see a non-zero coherence between the $\ket{20}_{01}$ and $\ket{02}_{01}$ elements. 
 An identical density operator $\rho_{2B}$ is obtained when we post-select on the events where two photons exit in fibers 2 and 3. 
 \par
To test this two-photon coherence, we connect the output fibers 0 and 1 of the DEMUX to a $2 \times 2$ beam splitter (BS)  (as shown in Fig. \ref{fig:setup}), and then connect the output fibers of the BS to detectors 0 and 1.  If we assume that the relative phase between the two fibers before the BS is $\phi$, and look at coincidence detections after the BS,  the detection probability is 
\begin{equation}
P_{2A}(\phi) = \frac{3}{4} + \frac{1}{4} \cos (2 \phi). 
\end{equation}
We thus predict oscillations with a frequency ($=2$) that is double that of single-photon interference, corresponding to self-interference of a two-photon wave packet with wavelength $\lambda/2$ \cite{rarity90,fonseca99b}.  The maximum expected visibility is given by $\mathcal{V}_2 = (3/4-1/4)/(3/4+1/4)=1/2$.  
 \par
\begin{figure}
\includegraphics[width=8cm]{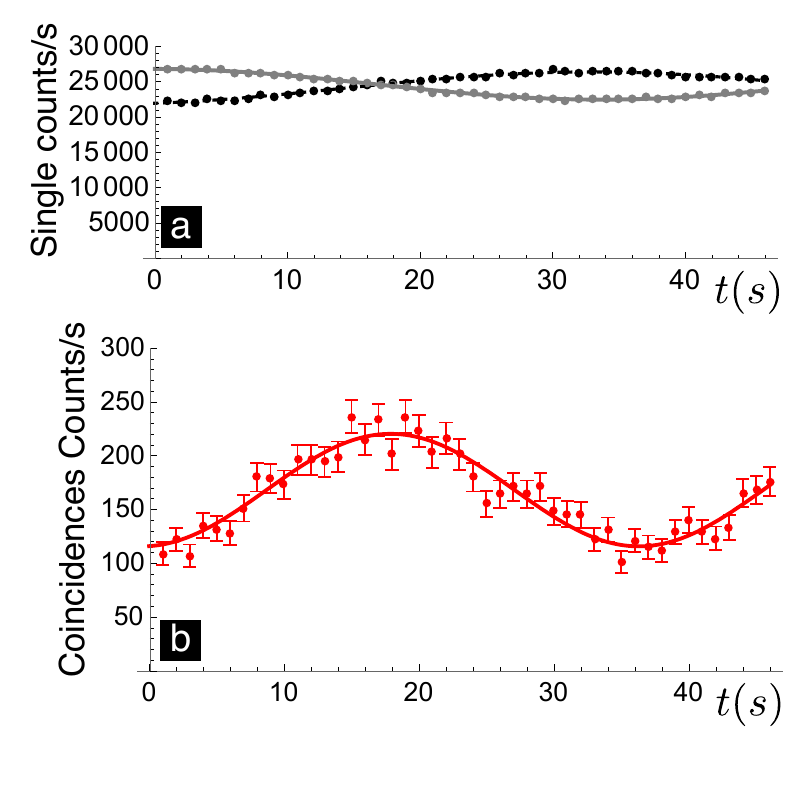}
\caption{Single and coincidence counts at detectors 0 and 1 when the phase in the corresponding $2\times2$ interferometer is varied over time.  The coincidence counts show the double frequency corresponding to self-interference of a two-photon wavepacket. Error bars correspond to Poissonian count statistics.}
\label{fig:TPI}
\end{figure}

 Figure \ref{fig:TPI}  shows experimental results for this configuration.  
 The phase was controlled using a telecomm phase modulator driven by an amplified function generator.   We registered single counts and coincidence counts as the phase was varied over time.  To compare the period of the two-photon interference with that of the single photon interference, we created a slight imbalance in the intensities of the WCSs, so that low contrast single-photon interference fringe could be observed.  Fitting the curves to a a sinusoidal function, we obtain single-photon periods $T_1=75\pm3$s  and $T^\prime_1=69\pm2$s, while the two-photon interference has period $T_2=36.8\pm0.8$s, confirming the 
 factor of two in the oscillation frequency and the self-interference of a two-photon wavepacket.   The visibility of the two photon interference was $\mathcal{V}_{2} \approx 0.52\pm0.3$, in agreement with the maximum predicted value, showing high-quality fourth-order interference. 
\section{Post-selected Bipartite State and Quantum Correlations}
\label{sec:discord}
To further our analysis, let us divide the output modes into two bipartitions. We post-select on events where one photon exits in partition $A$, composed of modes $0$ and $1$, while the other exits in partition $B$, composed of modes $2$ and $3$.  Let us switch from the multiple-rail notation to a qubit notation:  $\ket{1000} \equiv \ket{0}_A$, $\ket{0100} \equiv \ket{1}_A$, $\ket{0010} \equiv \ket{0}_B$ and $\ket{0001} \equiv \ket{1}_B$.  Following section \ref{sec:PSstate} and the total density operator given in \eqref{eq:rhotot}, the post-selected bipartite density operator is 
 \begin{equation}
\rho(\gamma)_{AB} =   \frac{1}{4}\left(
\begin{array}{cccc}
t & u & u & v \\
 u & w & w & u \\
 u & w & w & u \\
 v  & u & u &t \\
\end{array}
\right),
\label{eq:rhogen}
\end{equation}
where $t=(1+4\gamma+\gamma^2)/(1+ \gamma)^2$ , $u=(1-\gamma)/(1+ \gamma)$,  $v=(1-4\gamma+\gamma^2)/(1+ \gamma)^2$, $w=(1+\gamma^2)/(1+ \gamma)^2$.  Normalization requires $t+w=2$.    
\par
\begin{figure}
\includegraphics[width=8cm]{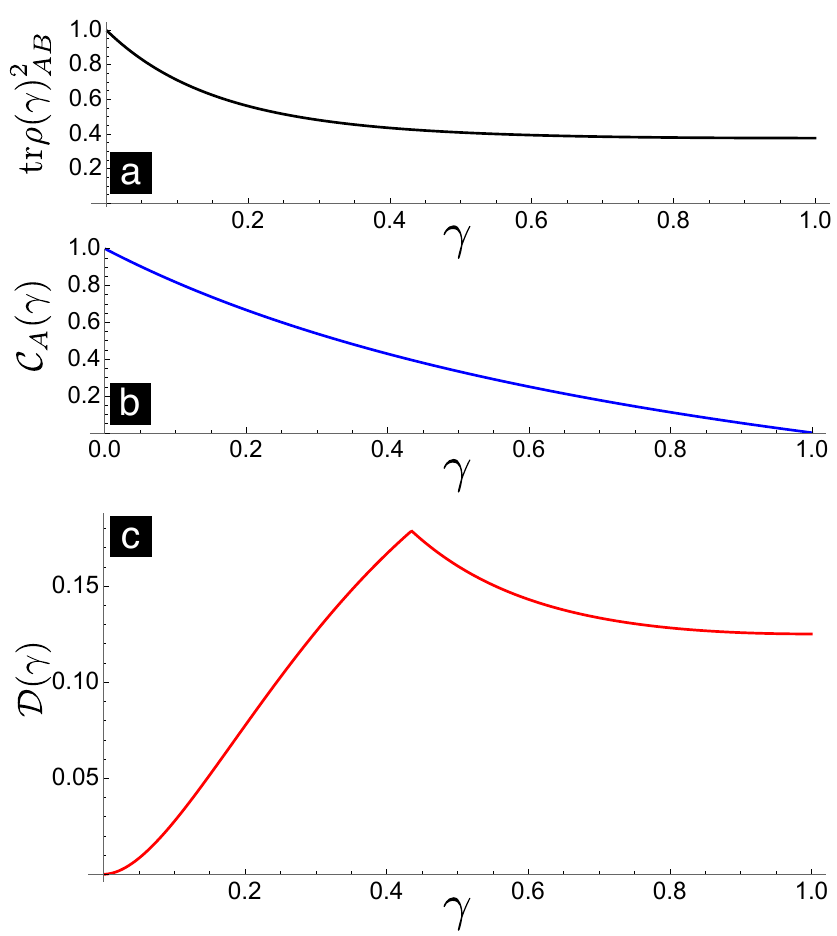}
\caption{Properties of the post-selected bipartite output state as a function of the intensity mismatch $\gamma$.  The bipartite purity a) and single-qubit coherence b) both decrease as $\gamma$ goes from 0 to 1.  The geometric discord c) is maximum for $\gamma \approx 0.435$, when the two input WCSs have unequal intensities.}
\label{fig:teo}
\end{figure}

The output state depends upon the ratio $\gamma$ between mean photon numbers in the input WCSs.  Without loss of generality, we consider the range $0 \leq \gamma \leq 1$, as $\gamma >1$ can be handled by simply interchanging $\eta$ and $\eta^\prime$. The purity, plotted in \ref{fig:teo} a), is given by $\mathrm{tr}\rho(\gamma)_{AB}^2 = (\gamma^4+4 \gamma^2 +1)/(1+\gamma)^4$, and reaches a minimum value of $3/8$ when $\gamma=1$.   In this case, $u=0$ and we have  
 \begin{equation}
\rho(1)_{AB} =  \frac{1}{8} \left(
\begin{array}{cccc}
3 & 0 & 0 & -1 \\
 0 & 1 & 1 & 0 \\
 0 & 1 & 1 & 0 \\
 -1  & 0 & 0 &3 \\
\end{array}
\right).
 \end{equation}

 This state is diagonal in the Bell-state basis
\begin{equation}
\rho(1)_{AB} =  \frac{1}{4} \left ( \ket{\Phi^+} \bra{\Phi^+}+2 \ket{\Phi^-} \bra{\Phi^-} +\ket{\Psi^+} \bra{\Psi^+}\right),
\end{equation}
and an example of a so-called ``X" state, with maximally mixed marginal density matrices $\rho_A=\rho_B=\mathbb{I}/2$ \cite{ali10,quesada12}.
The limit $\gamma=0$,  corresponding to a single WCS input, gives the post-selected pure state $\rho(0)_{AB}=\ket{+}\bra{+}$, where $\ket{\pm}=(\ket{0}\pm\ket{1})/\sqrt{2}$. 
\par
In the general case, the post-selected output state $\rho(\gamma)_{AB}$, though separable, displays quantum correlations.
To analyze this, it is useful to put the output state in its Bloch representation
\begin{align}
\rho(\gamma)_{AB} =&  \frac{1}{4} \left(I_2 \otimes I_2+ u \sigma_x \otimes I  + u I \otimes \sigma_x  \right. \nonumber \\ 
& \left. + \sum_{j} C_{jj} \sigma_j \otimes \sigma_j \right)
 \end{align}
where $I_2$ is the $2 \times 2$ identity matrix and $\sigma_i$ are the usual Pauli matrices with $i=x,y,z$. The state is characterized by real Bloch vectors $\vec{r}_A=\vec{r}_B=(u, 0 ,0)$ describing the local density operators, and the correlation matrix with real elements $C_{ij}=\langle \sigma_i \otimes\sigma_j\rangle$.  In our case, $C=\mathrm{diag}(w+v,w-v,t-w)/2$ is a diagonal matrix.  The reduced density matrices for qubits $A$ and $B$ are given by $\rho_A=\rho_B=(I_2+ u \sigma_x)/2$.  The single qubit coherence, $\mathcal{C}_A(\gamma)=2 |\langle \sigma_+ \rangle|=u$, is plotted in Fig. \ref{fig:teo} b) and is maximum for $\gamma=1$ and vanishes for $\gamma=0$. 
To evaluate quantum correlations, we use the geometric discord $\mathcal{D}$, following the recipe introduced in  Ref. \cite{dakic10}.   We find
\begin{align}
\mathcal{D}(\gamma) & = \frac{1}{2}\left(\frac{w^2+v^2}{2} + \frac{(t-w)^2}{4}+u^2 \right.  \nonumber \\
& \left. - \max\left [\frac{(w-v)^2}{4}, \frac{(t-w)^2}{4},\frac{(w+v)^2+4u^2}{4} \right] \right), 
\label{eq:GD}
\end{align}
where we choose a scaling parameter so that $\mathcal{D}=1$ for a maximally-entangled Bell state.
Using the definitions of $t,w,v$ just after Eq. \eqref{eq:rhogen}, we see that $(t-w)^2=(w-v)^2$, so that the maximum is between two quantities that depend on $\gamma$.  
A plot of the discord \eqref{eq:GD} for the state \eqref{eq:rhogen} is shown in Fig. \ref{fig:teo} c).   One can see a ``kink" at $\gamma \approx 0.435$, where the maximum value $\mathcal{D}(0.435)\sim0.178$ is obtained.  It is at this point that there is a change in the maximum function in Eq. \eqref{eq:GD}.   Similar kinks, corresponding to sudden changes of discord, have been studied in the context of quantum dynamics \cite{maziero09,fanchini10,shi10,auccaise11}, quantum phase transitions \cite{werlang10,maziero10},  and the quantum-classical transition \cite{mazzola10,cornelio12}.  For a more comprehensive survey of this topic, see the recent reviews  \cite{modi12, bera17}.     
In the present case, the sudden change is not related to evolution of the system but rather to the initial conditions at the source.   Thus, the quantum correlation can be controlled and maximized as a function of the intensity mismatch ratio of the input WCSs.   The case $\gamma=1$ as has been studied previously \cite{choi17} gives  $\mathcal{D}(1)=0.125$.  Thus, by using unbalanced intensities so that  $\gamma \sim 0.435$, the geometric discord $\mathcal{D}$ can be increased as compared to the the balanced case. 
\subsection{Experimental Evaluation of Quantum Correlations}
As discussed in the last section, we divide our experimental system into parts $A$ and $B$, each consisting of a spatial qubit spanned by the states corresponding to two output cores of the 4CF.  To realize projective measurements, the 4CF-BS is connected to a DEMUX and SMFs.  These are connected to phase modulators, and then attached to $2 \times 2$ beam splitters, respecting the bipartition (see Fig. \ref{fig:setup} and definition above).  Setting the relative phase between the two SMFs allows us to perform projective measurements onto local bases of the form $\ket{\phi}=(\ket{0} \pm \exp(i \phi) \ket{1})/\sqrt{2}$. For example, for $\phi=0$ we can project onto the  ${\sigma_x}$ eigenstates, while for  $\phi=\pi/2$ we project onto the eigenstates of ${\sigma_y}$.  The phase is controlled using telecomm phase modulators controlled by an FPGA and function generator, similar to the method reported in Ref. \cite{carine20}.   The elements of  the correlation matrix and Bloch vectors were estimated using the procedure described in \cite{sup}.     
\par
\begin{figure}
\includegraphics[width=8cm]{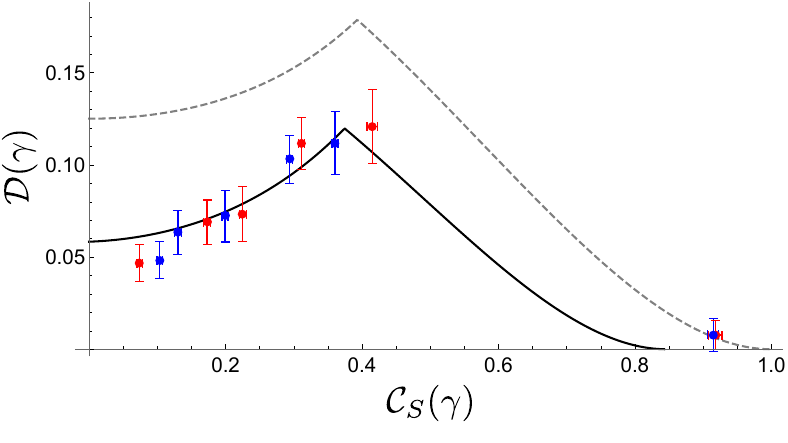}
\caption{Geometric discord $\mathcal{D}$ for the post-selected bipartite output state as a function of the local coherence $\mathcal{C}_S$ ($S=A,B$). The dashed grey upper curve is the theoretical prediction from \eqref{eq:GD}. The red and blue points correspond to experimental points obtained by considering the local marginals of $A$ or $B$, respectively.  The black curve is a the theoretical model taking into account dephasing and polarization mode mismatch.}
\label{fig:GD}
\end{figure}

Our measurement setup allows us to easily estimate local coherences simultaneously with joint projective measurements.  Following the theoretical model \eqref{eq:rhogen}, these are monotonic functions of $\gamma$.  Thus, for convenience, we experimentally estimated the geometric discord $\mathcal{D}$ as a function of the local coherence $\mathcal{C}_S$ ($S=A,B$).  Figure \ref{fig:GD} shows our experimental results.  The grey dashed line corresponds to our theoretical model, combining figures \ref{fig:teo} b) and c). Red and blue points correspond to discord calculated using the local marginals for systems $A$ and $B$, respectively.   These do not overlap with the ideal theoretical model, which we believe is due to experimental imperfections such as  phase fluctuations and polarization mode mismatch.  To test this, we consider the initial state \eqref{eq:rhogen} generated with imperfect mode overlap (using $\mathcal{V}_{WCS}$ from section \ref{sec:hom}), and phase damping. The model,  described in more detail in \cite{sup}, corresponds to the black solid line in Fig. \ref{fig:GD}, and is compatible with the experimental data.   The experimental points show that  the quantum correlations can be increased by manipulated the intensity mismatch of the input WCSs.      
 \section{Conclusions}
 We presented results demonstrating fourth-order interference of mutually incoherent classical laser pulses at a novel multi-port beam splitter device, imbedded within a multi-core optical fiber.  We observed photon/photon interference with visibility compatible with the ideal theoretical value.  In addition, the self-interference of the output two-photon wavepackets from photon coalescence was also tested, also giving high-quality visibility.  As a novel application of fourth-order interference in this type of multi-port device, we studied post-selected quantum correlations in the form of quantum discord.  We showed that the geometric discord can be maximized by controlling the intensity mismatch ratio between the input weak coherent laser pulses.  We expect these results to be useful in future quantum communications systems constructed within a space-division multiplexing infrastructure.   
\begin{acknowledgements}
 We thank G. Xavier for valuable conversations, and N. Guerrero and T. Garc\'ia for lab assistance. This work was  supported by Fondo Nacional de Desarrollo Cient\'{i}fico y Tecnol\'{o}gico (ANID) (11201348, 1200266, 1200859) and  ANID – Millennium Science Initiative Program – ICN17\_012.  JC was supported by ANID/REC/PAI77190088. 
 \end{acknowledgements}
 
\section{appendix: Full output density matrix}
The full density matrix containing all two-photon components can be
written in terms of the states (modes in vacuum state are omitted here) $\ket{2}_0$,$\ket{2}_1$,$\ket{2}_2$,$\ket{2}_3$,$\ket{11}_{01}$, $\ket{11}_{02}$, $\ket{11}_{03}$, $\ket{11}_{12}$, $\ket{11}_{13}$, $\ket{11}_{23}$.  Explicitly, it is 
\begin{widetext}

\begin{equation}
\frac{1}{16}\left(
\begin{array}{cccccccccc}
 \frac{a}{2} & \frac{c}{2} & \frac{a}{2} & \frac{c}{2} & \frac{b}{\sqrt{2}} & -\frac{a}{\sqrt{2}} &  -\frac{b}{\sqrt{2}} &  -\frac{b}{\sqrt{2}} & -\frac{c}{\sqrt{2}} & \frac{b}{\sqrt{2}} \\
 \frac{c}{2} & \frac{a}{2} & \frac{c}{2} & \frac{a}{2} & \frac{b}{\sqrt{2}} & -\frac{c}{\sqrt{2}} &  -\frac{b}{\sqrt{2}} &  -\frac{b}{\sqrt{2}} & -\frac{a}{\sqrt{2}} & \frac{b}{\sqrt{2}} \\
 \frac{a}{2} & \frac{c}{2} & \frac{a}{2} & \frac{c}{2} & \frac{b}{\sqrt{2}} & -\frac{a}{\sqrt{2}} &  -\frac{b}{\sqrt{2}} &  -\frac{b}{\sqrt{2}} & -\frac{c}{\sqrt{2}} & \frac{b}{\sqrt{2}} \\
 \frac{c}{2} & \frac{a}{2} & \frac{c}{2} & \frac{a}{2} & \frac{b}{\sqrt{2}} & -\frac{c}{\sqrt{2}} &  -\frac{b}{\sqrt{2}} &  -\frac{b}{\sqrt{2}} & -\frac{a}{\sqrt{2}} & \frac{b}{\sqrt{2}} \\
 \frac{b}{\sqrt{2}} & \frac{b}{\sqrt{2}} & \frac{b}{\sqrt{2}} & \frac{b}{\sqrt{2}} & d & -b & -d & -d & -b & d \\
 -\frac{d}{\sqrt{2}} & -\frac{c}{\sqrt{2}} & -\frac{a}{\sqrt{2}} & -\frac{c}{\sqrt{2}} & -b & a & b & b & c & -b \\
 -\frac{b}{\sqrt{2}} &  -\frac{b}{\sqrt{2}} &  -\frac{b}{\sqrt{2}} &  -\frac{b}{\sqrt{2}} & -d & b & d & d & b & -d \\
 -\frac{b}{\sqrt{2}} &  -\frac{b}{\sqrt{2}} &  -\frac{b}{\sqrt{2}} &  -\frac{b}{\sqrt{2}} & -d & b & d & d & b & -d \\
- \frac{c}{\sqrt{2}} & -\frac{a}{\sqrt{2}} & -\frac{c}{\sqrt{2}} & -\frac{a}{\sqrt{2}} & -b & c & b & b & a & -b \\
\frac{b}{\sqrt{2}} & \frac{b}{\sqrt{2}} & \frac{b}{\sqrt{2}} & \frac{b}{\sqrt{2}} & d & -b & -d & -d & -b & d \\
\end{array}
\right)
\label{eq:rhotot}
\end{equation}
where $a=(2+8\gamma+2\gamma^2)/M$, $b=(2-2\gamma^2)/M$, $c=(2-8\gamma+2\gamma^2)/M$, $d=(2+2\gamma^2)/M$ and $M=1+2 \gamma +  \gamma^2$. 
\par
When the input WCSs have equal intensity so that $\gamma=1$, we have
\begin{equation}
\label {eq:rhototequal}
\frac{1}{16}\left(
\begin{array}{cccccccccc}
 \frac{3}{2} & -\frac{1}{2} & \frac{3}{2} & -\frac{1}{2} & 0 & -\frac{3}{\sqrt{2}} & 0 & 0 & \frac{1}{\sqrt{2}} & 0 \\
 -\frac{1}{2} & \frac{3}{2} & -\frac{1}{2} & \frac{3}{2} & 0 & \frac{1}{\sqrt{2}} & 0 & 0 & -\frac{3}{\sqrt{2}} & 0 \\
 \frac{3}{2} & -\frac{1}{2} & \frac{3}{2} & -\frac{1}{2} & 0 & -\frac{3}{\sqrt{2}} & 0 & 0 & \frac{1}{\sqrt{2}} & 0 \\
 -\frac{1}{2} & \frac{3}{2} & -\frac{1}{2} & \frac{3}{2} & 0 & \frac{1}{\sqrt{2}} & 0 & 0 & -\frac{3}{\sqrt{2}} & 0 \\
 0 & 0 & 0 & 0 & 1 & 0 & -1 & -1 & 0 & 1 \\
 -\frac{3}{\sqrt{2}} & \frac{1}{\sqrt{2}} & -\frac{3}{\sqrt{2}} & \frac{1}{\sqrt{2}} & 0 & 3 & 0 & 0 & -1 & 0 \\
 0 & 0 & 0 & 0 & -1 & 0 & 1 & 1 & 0 & -1 \\
 0 & 0 & 0 & 0 & -1 & 0 & 1 & 1 & 0 & -1 \\
 \frac{1}{\sqrt{2}} & -\frac{3}{\sqrt{2}} & \frac{1}{\sqrt{2}} & -\frac{3}{\sqrt{2}} & 0 & -1 & 0 & 0 & 3 & 0 \\
 0 & 0 & 0 & 0 & 1 & 0 & -1 & -1 & 0 & 1 \\
\end{array}
\right)
\end{equation}
\end{widetext}


\end{document}